\documentclass[%
 reprint,
superscriptaddress,
 amsmath,amssymb,
 prl,
 aps,
 floatfix,
]{revtex4-1}

\usepackage{graphicx}
\usepackage{dcolumn}
\usepackage{bm}

\begin{document}

\preprint{http://arxiv.org/abs/1909.01204}

\title{Two-Dimensional Antimony Oxide}

\author{Stefan Wolff}
\email{stefan.wolff@fau.de}
\affiliation
{Department of Physics, Chair of Experimental Physics, Friedrich-Alexander-Universit\"at Erlangen-N\"urnberg (FAU), Staudtstr. 7, 91058 Erlangen, Germany}

\author{Roland Gillen}
\affiliation
{Department of Physics, Chair of Experimental Physics, Friedrich-Alexander-Universit\"at Erlangen-N\"urnberg (FAU), Staudtstr. 7, 91058 Erlangen, Germany}

\author{Mhamed Assebban}
\affiliation
{Instituto de Ciencia Molecular (ICMol), Universidad de Valencia, Catedr{\'a}tico Jos{\'e} Beltr{\'a}n 2, 46980, Paterna, Valencia, Spain.}
\affiliation
{Department of Chemistry and Pharmacy \& Joint Institute of Advanced Materials and Processes (ZMP), Friedrich-Alexander-Universit\"at Erlangen-N\"urnberg (FAU), Dr.-Mack-Stra{\ss}e 81, 90762, F\"urth, Germany.}

\author{Gonzalo Abell{\'a}n}
\affiliation
{Instituto de Ciencia Molecular (ICMol), Universidad de Valencia, Catedr{\'a}tico Jos{\'e} Beltr{\'a}n 2, 46980, Paterna, Valencia, Spain.}
\affiliation
{Department of Chemistry and Pharmacy \& Joint Institute of Advanced Materials and Processes (ZMP), Friedrich-Alexander-Universit\"at Erlangen-N\"urnberg (FAU), Dr.-Mack-Stra{\ss}e 81, 90762, F\"urth, Germany.}

\author{Janina Maultzsch}
\affiliation
{Department of Physics, Chair of Experimental Physics, Friedrich-Alexander-Universit\"at Erlangen-N\"urnberg (FAU), Staudtstr. 7, 91058 Erlangen, Germany}

\date{\today}

\begin{abstract}
Two-dimensional (2D) antimony, so-called antimonene, can form antimonene oxide when exposed to air. We present different types of single- and few-layer antimony oxide structures, based on density functional theory (DFT) calculations. Depending on stoichiometry and bonding type, these novel 2D layers have different structural stability and electronic properties, ranging from topological insulators to semiconductors with direct and indirect band gaps between 2.0 and 4.9\,eV. We discuss their vibrational properties and Raman spectra for experimental identification of the predicted structures.
\end{abstract}


\maketitle

The exfoliation of a single layer of graphene from bulk graphite unleashed a new field in physics and chemistry focusing on the investigation of two-dimensional (2D) layered crystals~\cite{Novoselov2004, Geim2007}. Over the past years an increasing number of 2D materials with vastly different properties have been discovered. Group-15 elements, also known as pnictogens, are suitable to form monoelemental 2D layered materials, which are promising candidates for a variety of applications in the field of plasmonics~\cite{Slotman2018}, for sensing~\cite{Mayorga-Martinez2019}, electronic~\cite{Pizzi2016, Chen2018, Zhang2019}, and optoelectronic~\cite{Wang2018} devices. Antimony is one of these elements and can form layered structures called antimonene. Recently, few-layer antimonene was realized experimentally by different methods such as epitaxial growth~\cite{Fortin-Deschenes2017, Wu2017, Chen2018Epi} or exfoliation~\cite{Ares2016, Gibaja2016, Ares2018, Lloret2019}. The electronic and vibrational properties of monolayer antimonene have been investigated theoretically, with a predicted band gap of about 2.4\,eV~\cite{Akturk2015, Gupta2015, Wang2015, Zhang2015, Gibaja2016, Ji2016, Wang2017, Zhang2017}. Antimonene also appears reactive to air, however, in contrast to black phosphorus~\cite{Liu2014, Favron2015, Li2016, Abellan2017JACS, Meng2017, Kistanov2018, Zhang2018}, it seems to form new stable structures after oxidation. An oxidation process may even be favorable for tailoring the electronic properties since the electronic band structure depends on the degree of oxidation. Additionally, antimonene seems to be the first elemental 2D material that forms a stable 2D oxide naturally. Because of its similarities to other 2D pnictogens, such processes may also occur for related materials. Oxidation could then be used to further increase the quality of such monoelemental materials by, e.g., their encapsulation between oxidized layers.

However, the actual structure of oxidized antimonene layers and their physical properties are unknown~\cite{Abellan2017}. In \mbox{Ref.~\cite{Zhang2017}}, for instance, the theoretical predictions are based on a monolayer antimonene with \mbox{Sb=O} double bonds perpendicular to the antimonene plane. Taking into account that bulk antimony oxides exist in several different compositions (\mbox{$\alpha$-Sb$_2$O$_3$}, \mbox{$\beta$-Sb$_2$O$_3$}, \mbox{Sb$_2$O$_5$}, and mixtures thereof)~\cite{Allen2013, Cody1979} and display polymorphism, other structures may exist, where the oxygen atoms are bound to at least two Sb atoms and are incorporated into the antimonene planes. Few-layer antimonene prepared by exfoliation is likely to undergo oxidation~\cite{Ares2016, Gibaja2016, Abellan2017, Ares2018, Lloret2019}, but knowledge about oxidized few-layer Sb is missing. As electronic properties of oxidized antimonene are expected to depend crucially on the structure and the bonding between oxygen and antimony atoms, a precise knowledge of the atomic structure is essential for developing this new material. The ability to control the oxidation process can then be used to tailor the electronic band structure.

In this Letter we present two-dimensional antimony oxide single- and few-layer structures with properties depending on bonding type and stoichiometry. Based on density functional theory (DFT) calculations, we show that our proposed novel 2D antimony oxide structures are semiconducting with direct and indirect band gaps between 2.0 and 4.9\,eV. Furthermore, we present their vibrational modes for experimental identification. We expect that semimetallic few-layer antimonene can naturally form heterostructures with semiconducting oxidized layers.

\begin{figure}[t!]
    \centering
    \includegraphics[width=\columnwidth]{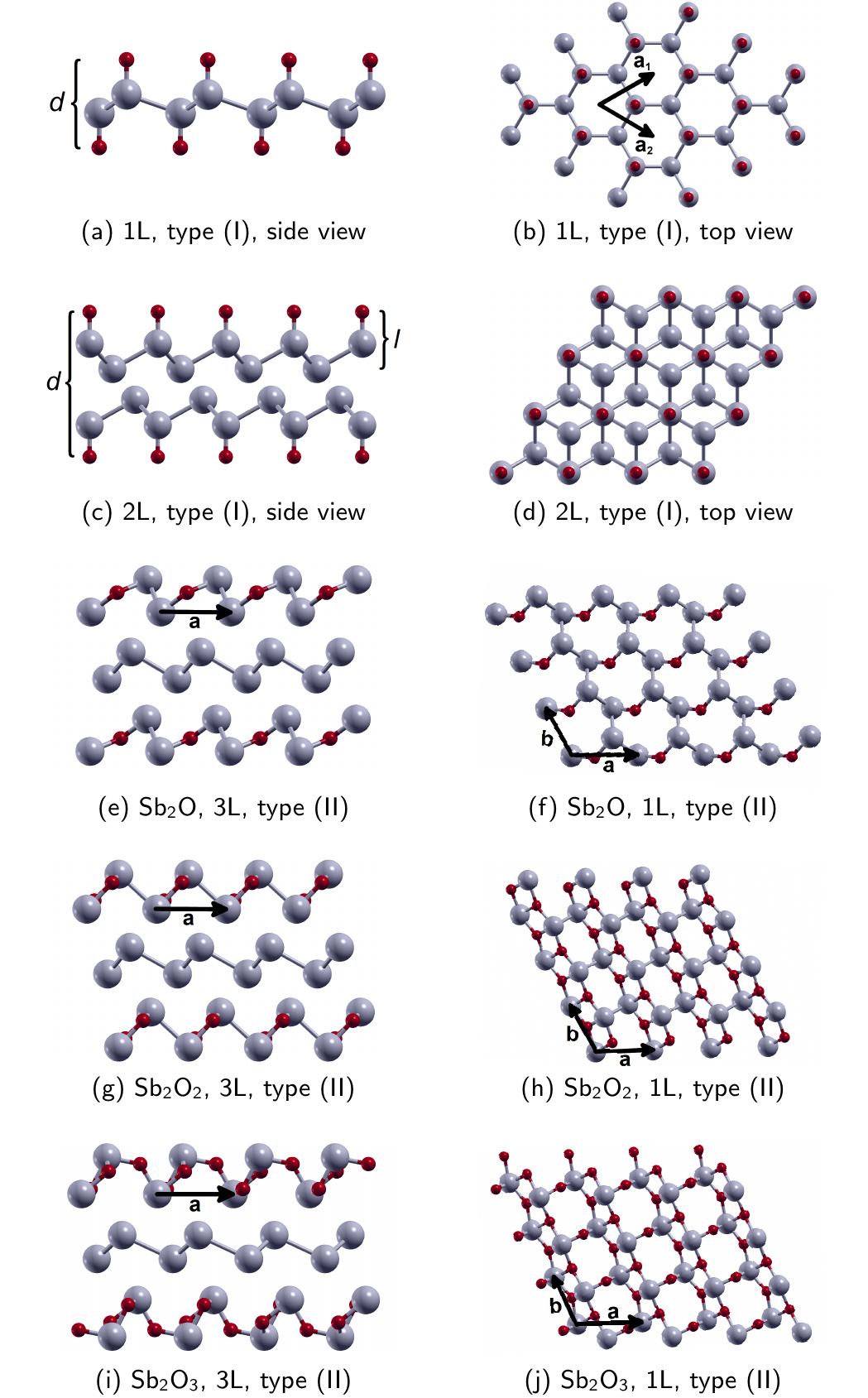}
    \caption{Type (I) antimonene oxide structures with one layer in side view (a) and top view (b), and two layers in (c) and (d). Type (II) antimonene oxide heterostructures with different stoichiometry of the oxidized layers: (e) and (f) Sb$_2$O; (g) and (h) Sb$_2$O$_2$; and (i) and (j) Sb$_2$O$_3$. Type (II) structures in top view show the oxidized layer only. Oxygen (antimony) atoms are shown in red (gray). Labels a and b on the figures indicate the in-plane lattice vectors. For details about the structural parameters, see Tables~S1 and S2.}
    \label{fig:structure}
\end{figure}

Antimony oxide mono- (1L) and bilayer (2L) structures with Sb=O double bonds, perpendicular to the plane, inspired by the fully oxidized antimonene monolayer proposed in \mbox{Ref.~\cite{Zhang2017}}, are shown in Figs.~\ref{fig:structure}(a)-\ref{fig:structure}(d). These structures are here referred to as \textit{type (I)}. Our simulations of the 1L phonon spectrum (see below and Fig.~S12) strongly suggests that the type (I) structures are metastable and likely transition into another more stable configuration. No stable system with more than two layers is found.

We introduce more stable antimonene oxide structures with different stoichiometries, where the oxygen atom is bound to at least two antimony atoms (Sb-O-Sb). We call these structures \textit{type (II)}, see Figs.~\ref{fig:structure}(e)-\ref{fig:structure}(j). They are obtained by a frozen-phonon approach by displacing the atoms in a unit cell of type (I) antimonene oxide according to a calculated phonon mode with negative frequency found in trilayer type (I) antimonene oxide. Using this approach on a trilayer system consisting of a non-oxidized antimonene layer, sandwiched by the bilayer shown in Fig.~\ref{fig:structure}(c), results in the structure of reduced symmetry shown in Figs.~\ref{fig:structure}(e) and \ref{fig:structure}(f). Corresponding to the unit cell of the top/bottom layer, we call this structure Sb$_2$O.

By increasing the amount of oxygen in the outer layers, such that the number of oxygen atoms matches the number of antimony atoms (Sb$_2$O$_2$), the structure changes even further: ``chains'' of alternating oxygen and antimony atoms are formed, which are each connected by three bonds to their neighbouring atoms. The chains are connected to each other by an additional bond between two antimony atoms, see Figs.~\ref{fig:structure}(g) and \ref{fig:structure}(h).

We increase the oxygen concentration in the outer layers to three oxygen atoms per two antimony atoms (Sb$_2$O$_3$), see Figs.~\ref{fig:structure}(i) and \ref{fig:structure}(j). The chain structure of alternating oxygen and antimony atoms is maintained. The additional oxygen atom is replacing the Sb-Sb bond, forming an Sb-O-Sb bond. As indicated in Fig.~\ref{fig:structure}, the different structures are labeled by the amount of antimony and oxygen atoms per unit cell in a single outer layer.

In few-layer antimonene the interlayer bonds have a significant covalent contribution to the otherwise noncovalent van der Waals interaction, as shown in previous experimental and theoretical work. Therefore we assume that the inner, nonoxidized layer will indeed be affected by a change of the lattice parameters induced by the oxidation of the outer layers. The trilayer structures shown in Figs.~\ref{fig:structure}(e), \ref{fig:structure}(g), \ref{fig:structure}(i) thus provide a qualitative structural picture in comparison to the original type (I) structures. However, the small unit cells used in our calculations combined with a significant change of the lattice vectors of the oxidized layers with increasing oxidation induce a strain of up to 17\% in the inner, nonoxidized layers. We refer to Sec. 5 of the Supplemental Material for a detailed discussion. We note that for the Sb$_2$O$_3$ monolayer, the structure depicted in Fig.~\ref{fig:structure}(j) is not fully dynamically stable and relaxes into a slightly distorted geometry if a larger unit cell is used (Fig.~S4). Because of the small energy difference (5\,meV per Sb$_2$O$_3$ formula unit), we will further use the idealized Sb$_2$O$_3$ layer shown in Figs.~\ref{fig:structure}(i) and \ref{fig:structure}(j) for reasons of convenience. For real samples, we expect the formation of an amorphouslike antimonene oxide capping layer.

We now turn to the properties of the individual antimonene oxide monolayers with different degrees of oxidation, depicted in Figs.~\ref{fig:structure}(f), \ref{fig:structure}(h), \ref{fig:structure}(i), and bilayers thereof. The atomic positions and lattice vectors of all isolated structures were optimized. The phase transformation from type (I) to type (II) lowers the total energies of 1L and 2L structures by roughly 3.9 and 2.7\,eV, respectively, indicating that type (I) structures are metastable at best and transform into type (II)-like arrangements. To further confirm this, we performed three sets of molecular dynamics (MD) simulations at a temperature of 300\,K. In the first set (Fig.~S5), we started with the hexagonal primitive cell of monolayer antimonene and added three oxygen atoms close to the antimonene layer. The resulting equilibrated structure closely resembles the Sb$_2$O$_3$ structure of Fig.~\ref{fig:structure}(j). In the second set (Fig.~S6), a $4\times4$ supercell of hexagonal antimonene was used and an oxygen molecule was added at a distance of around 3\,\AA\space to the surface roughly every 1\,ps. In the third set (Fig.~S7), we started with a $2\times2$ unit cell of type (I) antimonene oxide and observed the evolution of the atomic structure. The lattice vectors were allowed to relax in all calculations. In all three sets of calculations, the oxygen is incorporated into the antimonene layer and chains of Sb-O bonds are formed. The well-known bulk antimony oxides with Sb$_2$O$_3$ and Sb$_2$O$_5$ stoichiometries have no common structural motif with the 2D antimonene oxide layers presented here.

For an equal number of atoms in a given type (II) structure, the total energy decreases by roughly 1.3\,eV per oxygen atom for an increasing amount of oxygen. This has been verified by calculating the energy of an O$_2$ molecule and adding or subtracting the energy, respectively. The emergence of 2D antimonene oxide with a higher degree of oxidation than Sb$_2$O$_3$, however, is rather unlikely: Any further increase of the oxygen concentration did not result in stable 2D layers, since the number of antimony atoms which oxygen atoms can bind to is limited.

\begin{figure}[t!]
    \centering
    \includegraphics[width=\columnwidth]{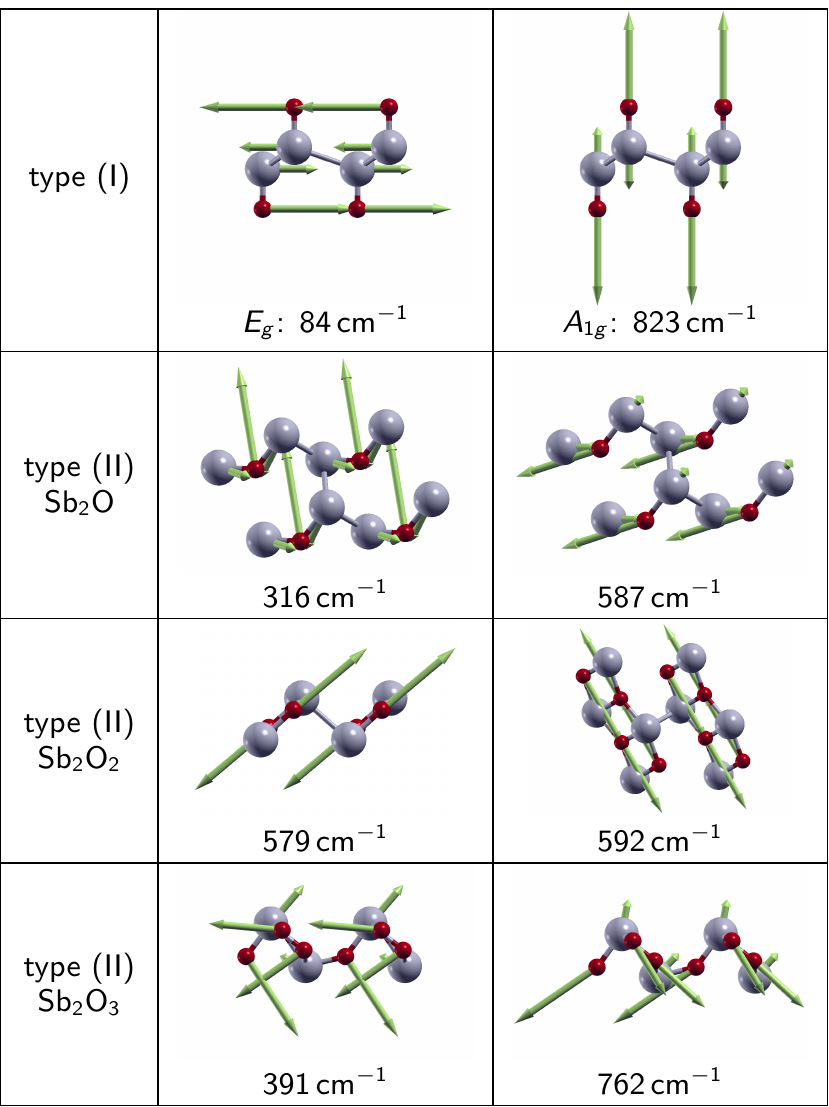}
    \caption{Exemplary display of phonon modes of the monolayer antimonene oxide structures investigated here. The type (II) Sb$_2$O and Sb$_2$O$_2$ on the left-hand side are shown from the top. All other structures are shown from the side. Green arrows indicate the displacements of the atoms and are to logarithmic scale.}
    \label{fig:phTable}
\end{figure}

In order to provide a guideline for identifying the different antimonene oxide structures experimentally, e.g., by Raman spectroscopy, we present their vibrational properties in the following.

\begin{figure}[t!]
    \centering
    \includegraphics[width=\columnwidth]{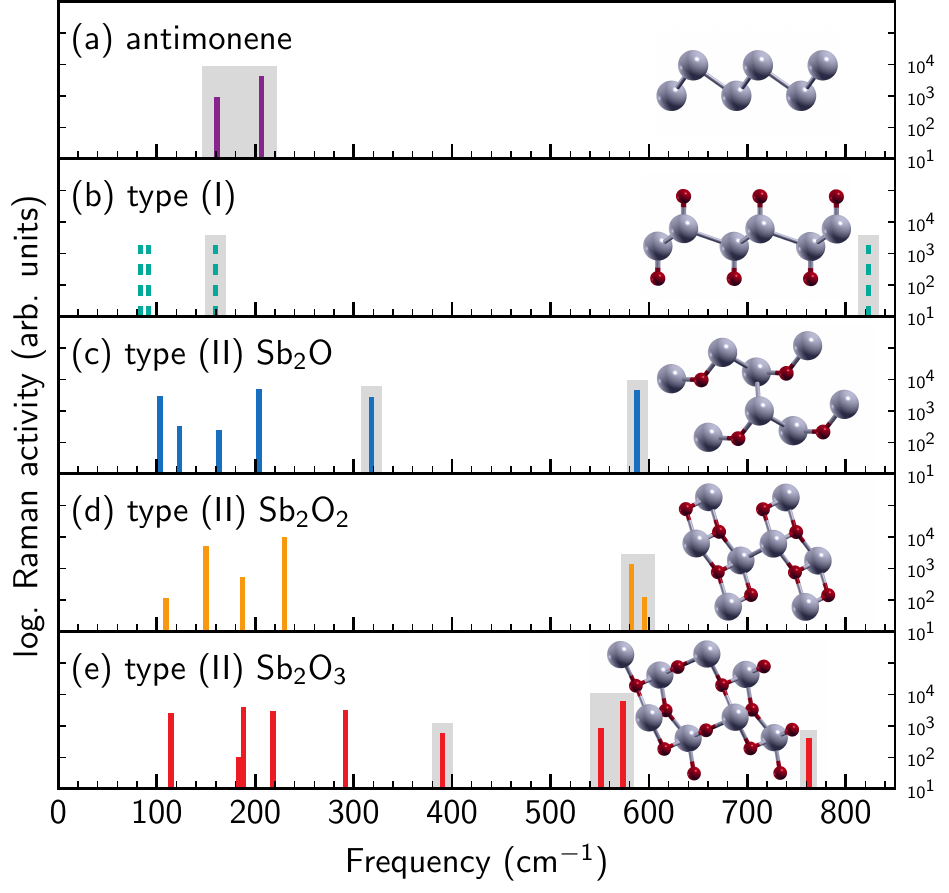}
    \caption{Calculated frequencies of Raman-active vibrational modes in (a) antimonene, (b) type (I), (c)-(e) type (II) antimonene oxide structures, as indicated in the insets and shown in {Fig.~\ref{fig:structure}}. Except for (b), the height of the bars indicates the calculated Raman activity (log. scale).}
    \label{fig:raman}
\end{figure}

The type (I) structures belong, like pristine antimonene, to the $D_{3d}$ symmetry group; therefore the vibrational modes include modes with $E_g$ and $A_{1g}$ symmetry (Fig.~\ref{fig:phTable}). All twelve modes of type (I) monolayers (four atoms per unit cell, see also Fig.~S8) fall into two regions, one below 180$\,\text{cm}^{-1}$ and one at around 820$\,\text{cm}^{-1}$, see Fig.~\ref{fig:raman}. The latter corresponds to stretching of the Sb=O bonds and is indicative of a type (I) structure. The frequency range below 180$\,\text{cm}^{-1}$ comprises the vibrations within the Sb layers; in addition there are rigid-layer vibrations in the case of the bilayer system.

The calculation of the phonon dispersion of the type (I) monolayer structure (Fig.~S12) results in negative frequencies of the acoustic branches over a large region of the Brillouin zone. This indicates that such structures are not stable experimentally.

The highest phonon frequency in the type (II) Sb$_2$O structure is at about $590\,\text{cm}^{-1}$ and is a mode with an out-of-plane component, see Fig.~\ref{fig:phTable}. A second characteristic mode of the type (II) Sb$_2$O layer, at about $315\,\text{cm}^{-1}$, is dominated by a motion of the oxygen atoms (Fig.~\ref{fig:phTable}~and~\ref{fig:raman}). For the displacement patterns of all vibrational modes, see Fig.~{S9}. The symmetry of these structures is $C_1$ for 1L and $C_i$ for 2L and 3L.

The monolayer type (II) Sb$_2$O$_2$ structure [Figs.~\ref{fig:structure}(g) and \ref{fig:structure}(h)] can be further symmetrized such that it corresponds to the $C_\text{2h}$ symmetry group. The frequency of the out-of-plane mode decreases to about $530\,\text{cm}^{-1}$ and the mode becomes Raman inactive. Two additional modes arise at around $590\,\text{cm}^{-1}$, which are also dominated by a displacement of oxygen atoms. The highest-frequency mode is an in-plane vibration, whereas the other mode is along the bond between oxygen and antimony atoms perpendicular to the direction of the Sb-O-Sb chain (Figs.~\ref{fig:phTable} and \ref{fig:raman}); see Fig.~S10 for all displacement patterns.

The additional oxygen atom in the type (II) Sb$_2$O$_3$ structure leads to five atoms per unit cell for 1L, i.e., 15 phonon modes, which are illustrated in Fig.~S11. In comparison to the previously discussed structures, the additional vibrational modes occur at roughly $300\,\text{cm}^{-1}$, $750\,\text{cm}^{-1}$, and in the range of $375$ to $430\,\text{cm}^{-1}$ for different layer numbers (Figs.~\ref{fig:phTable} and \ref{fig:raman}).

The phonon modes observed in the monolayer type (II) structures also appear in the respective 2L and 3L structures. The displacement patterns of these modes are qualitatively maintained and show overall similar frequencies, though some are shifted by up to about $50\,\text{cm}^{-1}$.

The characteristic $E_{g}$ and $A_{1g}$ Raman modes of pristine monolayer antimonene at $168$ and 206$\,\text{cm}^{-1}$ are shown in Fig.~\ref{fig:raman}(a). In \mbox{Figs.~\ref{fig:raman}(b)-\ref{fig:raman}(e)}, the positions of the Raman active modes in the monolayer type (I) and type (II) structures with different oxygen concentration are presented. The characteristic frequencies shown in Fig.~\ref{fig:raman} can be used for experimental identification of different antimonene oxide structures.

Our predictions of stable type (II) antimonene oxide layers are in agreement with recent experiments on the oxidation behavior of liquid-phase exfoliated few-layer antimonene~\cite{Assebban2019}. \mbox{Reference~\cite{Assebban2019}} reports the formation of a passivation layer on the surface, which shows evidence for \mbox{Sb$_2$O$_3$-like} layers. Raman measurements reveal characteristic modes in the range of 190-$450\,\text{cm}^{-1}$. This experimentally rules out the formation of type (I) structures. Instead, predicted phonon modes of the Sb$_2$O$_3$ layers fit reasonably well to the experimentally observed spectra~\cite{Assebban2019}.

\begin{figure}[t!]
    \centering
    \includegraphics[width=1.0\columnwidth]{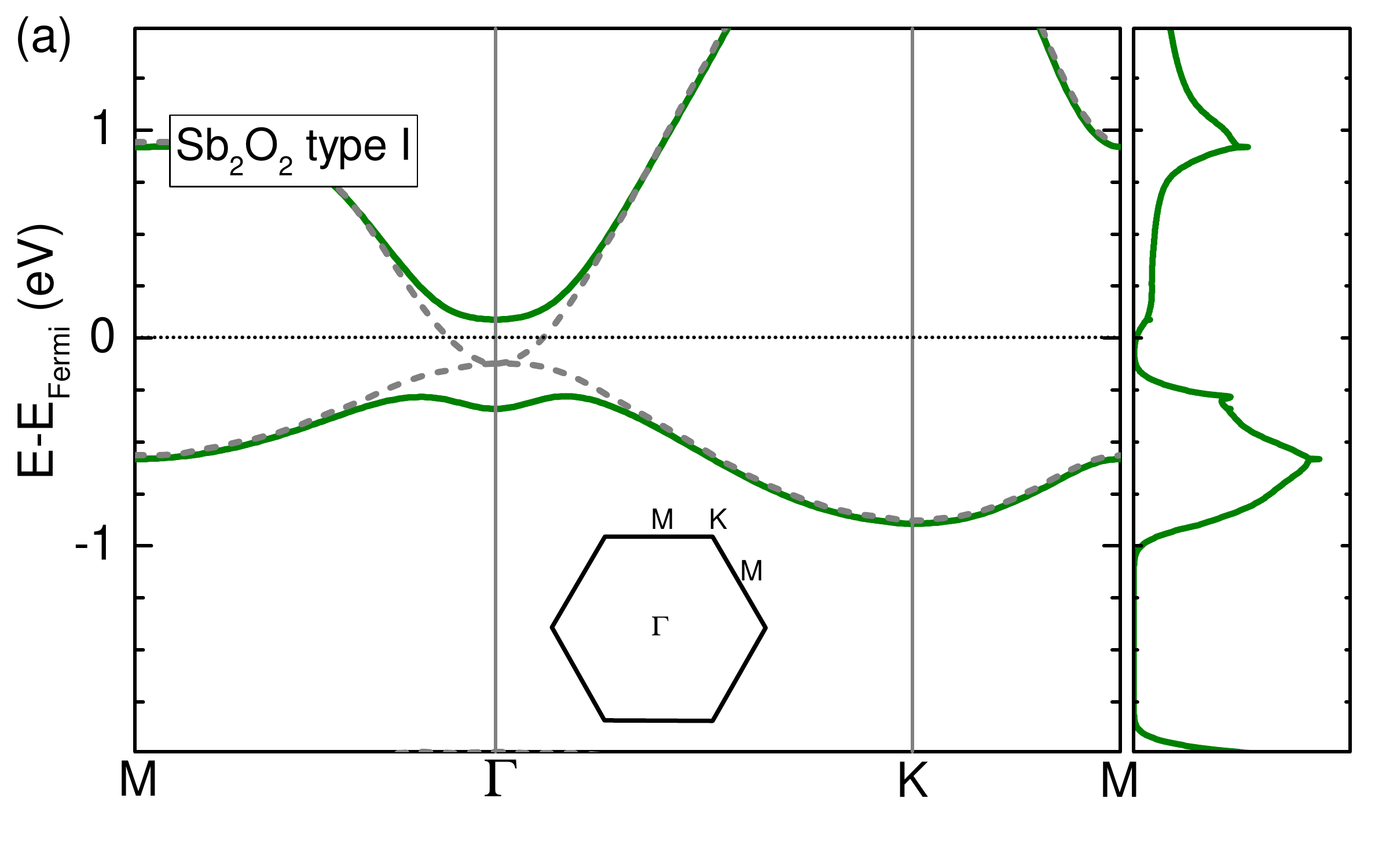}
    \includegraphics[width=1.0\columnwidth]{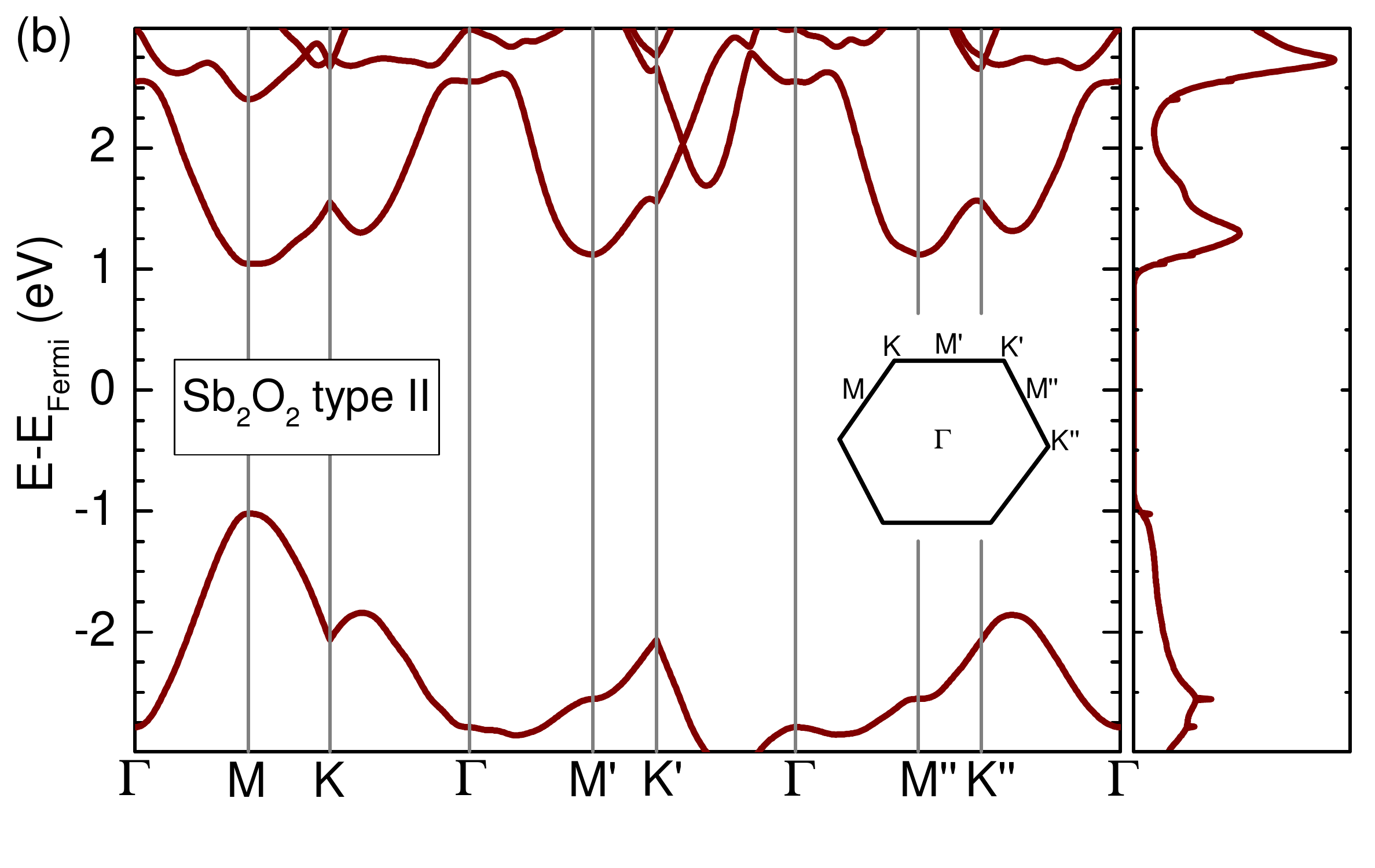}
    \includegraphics[width=1.0\columnwidth]{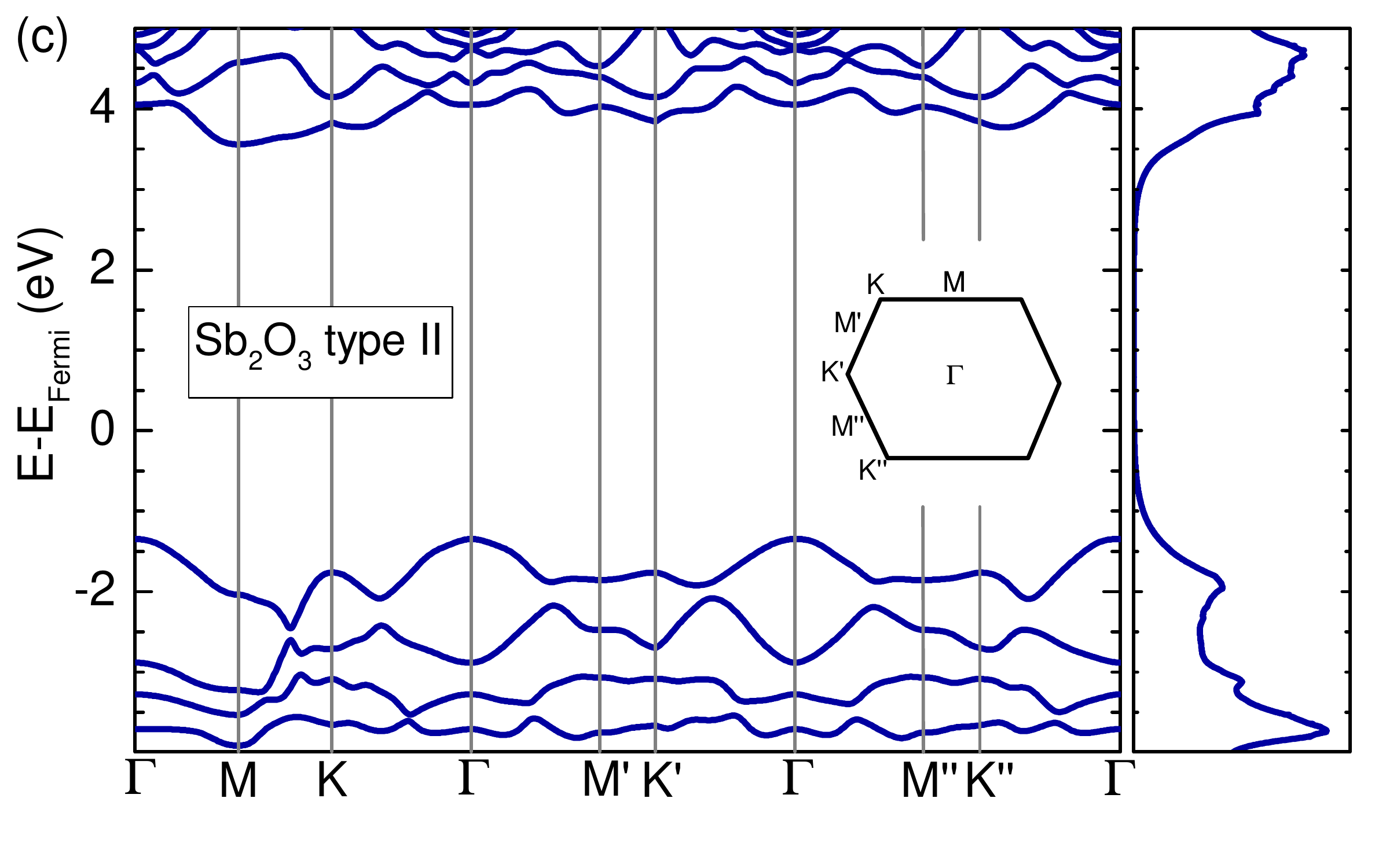}
    \caption{Electronic band structures and density of states calculated using the hybrid functional HSE12~\cite{hse12} and inclusion of spin-orbit interactions (SOI) for (a) type (I) Sb$_2$O$_2$, (b) type (II) Sb$_2$O$_2$, and (c) type (II) Sb$_2$O$_3$ monolayers. Gray dashed lines are the results without SOI and are almost congruent with the bands with SOI for (b) and (c). The zero of energy is set to the Fermi energy.}
    \label{fig:bands}
\end{figure}

While bulk and few-layer antimonene are metallic due to a partial covalent bonding between the layers, first-principles calculations on the GW level predict a value of about 2.4\,eV for nonoxidized monolayer antimonene~\cite{Wang2017}. Fully oxidized monolayer antimonene with double-bonded oxygen atoms [type (I)] was previously predicted to be a topological insulator with a small ``bulk'' band gap if spin-orbit interaction is included~\cite{Zhang2017}. Figure~\ref{fig:bands} shows the calculated electronic band structures of oxidized monolayer antimonene using the hybrid functional HSE12~\cite{hse12}, based on the atomic geometries of Fig.~\ref{fig:structure}(a) [type (I)] and our proposed structures [type (II)] from Figs.~\ref{fig:structure}(h) and \ref{fig:structure}(j). For type (I) Sb$_2$O$_2$, our calculations agree with the results of \mbox{Ref.~\cite{Zhang2017}}, showing a band gap of 168\,meV; the system is metallic if spin-orbit interaction is neglected [gray dashed lines in Fig.~\ref{fig:bands}(a)]. We refer to Sec. 5 of the Supplemental Material for a discussion of the electronic band structure of the trilayer structures shown in Figs.~\ref{fig:structure}(e), \ref{fig:structure}(g), \ref{fig:structure}(i). For type (II) Sb$_2$O$_2$, with the more stable chainlike configuration, our calculations predict the system to be a trivial insulator with a direct band gap of about 2.0\,eV at the edge of the Brillouin zone [Fig.~\ref{fig:bands}(b)]. In the latter case, there is no discernable effect of spin-orbit coupling on the electronic dispersion. Increasing the oxygen content in the unit cell [type (II) Sb$_2$O$_3$, Fig.~\ref{fig:structure}(j)] causes a transition from direct to indirect semiconductor and significantly increases the band gap to 4.9\,eV. This suggests that both the size and the nature of the fundamental band gap in oxidized antimonene could be tuned from the visible to the ultraviolet range, if control over the oxidation can be achieved. We expect a natural formation of heterostructures in which multiple semimetallic antimonene layers are sandwiched between semiconducting oxidized antimonene layers, due to the high reactivity between antimony and oxygen observed in \mbox{Ref.~\cite{Assebban2019}}.

In summary, we present layered antimonene oxide structures with the oxygen atoms incorporated into the antimonene sheet [type (II)]. They are more stable than configurations with Sb=O double bonds perpendicular to the antimonene plane [type (I)]. Distinct differences in the vibrational frequencies between type (I) and different type (II) antimonene oxides allow an experimental identification of the structures via Raman spectroscopy. This is in good agreement with recent experimental findings on liquid-phase exfoliated few-layer antimonene~\cite{Assebban2019}. All type (II) single-layer antimonene oxides presented here are semiconductors with stochiometry-dependent band gaps ranging from approximately 2.0 to 4.9\,eV. Our results thus pave the way for tailoring the electronic band structure of antimonene flakes via controlled oxidation and will guide future development of antimonene-based 2D materials and heterostructures.

\begin{acknowledgments}
Computational resources used for the calculations were provided by the HPC of the Regional Computer Centre Erlangen (RRZE). This work has been supported by the Deutsche Forschungsgemeinschaft (DFG) within the CRC 953 (B13), by the European Union (ERC-2018-StG 804110-2D-PnictoChem to G.A.), and by the Spanish MINECO (Structures of Excellence Mar{\'i}a de Maeztu MDM-2015-0538). G.A. acknowledges support by the Generalitat Valenciana (CIDEGENT/2018/001), and the DFG (FLAG-ERA AB694/2-1).
\end{acknowledgments}

See Supplemental Material at [URL will be inserted by publisher] for computational details, structural parameters, images of all structures, snapshots of molecular dynamic calculations, vibrational modes, displacement patterns, phonon dispersion relations of the monolayer structures, electronic band structures of few layer systems, and the atomic positions used for the phonon dispersion calculations. Refs. \cite{QE-2009, GARRITY2014446, VANSETTEN201839, SIESTA_2002} are cited in the Supplemental Material.

\end{document}